# An extensive analysis of the presence of altmetric data for Web of Science publications across subject fields and research topics

Zhichao Fang[1*], Rodrigo Costas[1,2], Wencan Tian[3], Xianwen Wang[3], Paul Wouters[1]

* Corresponding author

Zhichao Fang (ORCID: 0000-0002-3802-2227)
[1] Centre for Science and Technology Studies (CWTS), Leiden University, Leiden, The Netherlands.
E-mail: z.fang@cwts.leidenuniv.nl

Rodrigo Costas (ORCID: 0000-0002-7465-6462)
[1] Centre for Science and Technology Studies (CWTS), Leiden University, Leiden, The Netherlands.
[2] DST-NRF Centre of Excellence in Scientometrics and Science, Technology and Innovation Policy, Stellenbosch University, Stellenbosch, South Africa.
E-mail: rcostas@cwts.leidenuniv.nl

Wencan Tian (ORCID: 0000-0001-7420-9315)
[3] WISE Lab, Institute of Science of Science and S&T Management, Dalian University of Technology, Dalian, China.
E-mail: tianwen@mail.dlut.edu.cn

Xianwen Wang (ORCID: 0000-0002-7236-9267)
[3] WISE Lab, Institute of Science of Science and S&T Management, Dalian University of Technology, Dalian, China.
E-mail: xianwenwang@dlut.edu.cn

Paul Wouters (ORCID: 0000-0002-4324-5732)
[1] Centre for Science and Technology Studies (CWTS), Leiden University, Leiden, The Netherlands.
E-mail: p.f.wouters@cwts.leidenuniv.nl

**Abstract**

Sufficient data presence is one of the key preconditions for applying metrics in practice. Based on both Altmetric.com data and Mendeley data collected up to 2019, this paper presents a state-of-the-art analysis of the presence of 12 kinds of altmetric events for nearly 12.3 million Web of Science publications published between 2012 and 2018. Results show that even though an upward trend of data presence can be observed over time, except for Mendeley readers and Twitter mentions, the overall presence of most altmetric data is still low. The majority of altmetric events go to publications in the fields of Biomedical and Health Sciences, Social Sciences and Humanities, and Life and Earth Sciences. As to research topics, the level of attention received by research topics varies across altmetric data, and specific altmetric data show different preferences for research topics, on the basis of which a framework for identifying *hot* research topics is proposed and applied to detect research topics with higher levels of attention garnered on certain altmetric data source. Twitter mentions and policy document citations were selected as two examples to identify hot research topics of interest of Twitter users and policy-makers, respectively, shedding light on the potential of altmetric data in monitoring research trends of specific social attention.

**Introduction**

Ever since the term "altmetrics" was coined in Jason Priem's tweet in 2010,[1] a range of theoretical and practical investigations have been taking place in this emerging area (Sugimoto et al., 2017). Given that many types of altmetric data outperform traditional citation counts with regard to the accumulation speed after publication (Fang & Costas, 2020), initially, altmetrics were expected to serve as faster and more fine-grained alternatives to measure scholarly impact of research outputs (Priem et al., 2010, 2012). Nevertheless, except for Mendeley readership which was found to be moderately correlated with citations (Zahedi et al., 2014; Zahedi & Haustein, 2018), a series of studies have confirmed the negligible or weak correlations between citations and most altmetric indicators at the publication level (Bornmann, 2015b; Costas et al., 2015; de Winter, 2015; Zahedi et al., 2014), indicating that altmetrics might capture diverse forms of impact of scholarship which are different from citation impact (Wouters & Costas, 2012).

The diversity of impact beyond science reflected by altmetrics, which is summarized as "broadness" by Bornmann (2014) as one of the important characteristics of altmetrics, relies on diverse kinds of altmetric data sources. Altmetrics do not only include events on social and mainstream media platforms related to scholarly content or scholars, but also incorporate data sources outside the social and mainstream media ecosystem such as policy documents and peer review platforms (Haustein et al., 2016). The expansive landscape of altmetrics and their fundamental differences highlight the importance of keeping them as separate entities without mixing, and selecting datasets carefully when making generalizable claims about altmetrics (Alperin, 2015; Wouters et al., 2019). In this sense, data presence, as one of the significant preconditions for applying metrics in research evaluation, also needs to be analyzed separately for various altmetric data sources.

*Presence of altmetric data for scientific publications*

Bornmann (2016) regarded altmetrics as one of the hot topics in the field of Scientometrics for several reasons, being one of them that there are large altmetric data sets available to be empirically analyzed for studying the impact of publications. However, according to existing studies, there are important differences of data coverage across diverse altmetric data. In one of the first, Thelwall et al. (2013) conducted a comparison of the correlations between citations and 11 categories of altmetric indicators finding that, except for Twitter mentions, the coverage of all selected altmetric data of PubMed articles was substantially low. This observation was reinforced by other following studies, which provided more evidence about the exact coverage for Web of Science (WoS) publications. Based on altmetric data retrieved from ImpactStory (IS), Zahedi et al. (2014) reported the coverage of four types of altmetric data for a sample of WoS publications: Mendeley readers (62.6%), Twitter mentions (1.6%), Wikipedia citations (1.4%), and Delicious bookmarks (0.3%). In a follow up study using altmetric data from Altmetric.com, Costas et al. (2015) studied the coverage of five altmetric data for WoS publications: Twitter mentions (13.3%), Facebook mentions (2.5%), blogs citations (1.9%), Google+ mentions (0.6%), and news mentions (0.5%). They also found that research outputs in the fields of Biomedical and Health Sciences and Social Sciences and Humanities showed the highest altmetric data coverage in terms of these five altmetric data. Similarly, it was reported by Haustein et al. (2015) that the coverage of five social and mainstream media data for WoS papers varied as follows: Twitter mentions (21.5%), Facebook mentions (4.7%), blogs citations (1.9%), Google+ mentions (0.8%), and news mentions (0.7%).

In addition to aforementioned large-scale research on WoS publications, there have been also studies focusing on the coverage of altmetric data for research outputs from a certain subject field or publisher. For example, on the basis of selected journal articles in the field of Humanities, Hammarfelt (2014) investigated the coverage of five kinds of altmetric data, including Mendeley readers (61.3%), Twitter mentions (20.6%), CiteULike readers (5.2%), Facebook mentions (2.9%), and blogs citations (2.2%). Waltman and Costas (2014) found that just about 2% of the publications in the biomedical literature received at least one F1000Prime recommendation. For papers published in the Public Library of Science (PLoS) journals, Bornmann (2015a) reported the coverage of a group of altmetric data sources tracked by PLoS's Article-Level Metrics (ALM). Since the data coverage is a value usually computed for most altmetric studies, similar coverage levels are found scattered across many other studies as well (Alperin, 2015; Fenner, 2013; Robinson-García et al., 2014). By summing up the total number of publications and those covered by altmetric data in 25 related studies, Erdt et al. (2016) calculated the aggregated percentage of coverage for 11 altmetric data. Their aggregated results showed that Mendeley readers covers the highest share of publications (59.2%), followed by Twitter mentions (24.3%) and CiteULike readers (10.6%), while other altmetric data show relatively low coverage in general (below 10%).

---
[1] On September 29, 2010, Jason Priem posted a tweet with the hashtag "altmetrics". See more details about this tweet at: https://twitter.com/jasonpriem/status/25844968813

*Identification of hot research topics using altmetric data*

The distributions of publications and article-level metrics across research topics are often uneven, which has been observed through the lens of text-based (Gan & Wang, 2015), citation-based (Shibata et al., 2008), usage-based (Wang et al., 2013), and altmetric-based (Noyons, 2019) approaches, making it possible to identify research topics of interest in different contexts, namely, the identification of *hot research topics*. By combining the concept made by Tseng et al. (2009), hot research topics are defined as topics that are of particular interest to certain communities such as researchers, Twitter users, Wikipedia editors, policy-makers, etc. Thus, *hot* is defined as the description of a relatively high level of attention that research topics have received on different altmetric data sources. *Attention* here is understood as the amount of interactions that different communities have generated around research topics, therefore those topics with high levels of attention can be identified and characterized as hot research topics from an altmetric point of view.

Traditionally, several text-based and citation-based methodologies have been widely developed and employed in detecting research topics of particular interest to researchers, like co-word analysis (Ding & Chen, 2014; Lee, 2008), direct citation and co-citation analysis (Chen, 2006; Small, 2006; Small et al., 2014), and the "core documents" based on bibliographic coupling (Glänzel & Czerwon, 1996; Glänzel & Thijs, 2012), etc. Besides, usage metrics, which are generated by broader sets of users through various behaviors such as viewing, downloading, or clicking, have been also used to track and identify hot research topics. For example, based on the usage count data provided by Web of Science, Wang and Fang (2016) detected hot research topics in the field of Computational Neuroscience, which are listed as the keywords of the most frequently used publications. By monitoring the downloads of publications in *Scientometrics*, Wang et al. (2013) identified hot research topics in the field of Scientometrics, operationalized as the most downloaded publications in the field.

From the point of view that altmetrics can capture the attention around scholarly objects from broader publics (Crotty, 2014; Sugimoto, 2015), some altmetric data were also used to characterize research topics based on the interest exhibited by different altmetric and social media users. For example, Robinson-Garcia et al. (2019) studied the field of Microbiology to map research topics which are highly mentioned within news media outlets, policy briefs, and tweets over time. Zahedi and van Eck (2018) presented an overview of specific topics of interest of different types of Mendeley users, like professors, students, and librarians, and found that they show different preferences in reading publications from different topics. Fang and Costas (2020) identified research topics of publications that are faster to be mentioned by Twitter users or cited by Wikipedia page editors, respectively. By comparing the term network based on author keywords of climate change research papers, the term network of author keywords of those tweeted papers, and the network of "hashtags" attached to related tweets, Haunschild et al. (2019) concluded that Twitter users are more interested in topics about the consequences of climate change to humans, especially those papers forecasting effects of a changing climate on the environment.

*Objectives*

Although there are multiple previous studies discussing the coverage of different altmetric data, after nearly ten years of altmetric research, we find that a renewed large-scale empirical analysis of the up-to-date presence of altmetric data for WoS publications is highly relevant. Particularly, since amongst previous studies, there still exist several types of altmetric data sources that have not been quantitatively analyzed. Moreover, although the correlations between citations and altmetric indicators have been widely analyzed at the publication level in the past, the correlations of their presence at the research topic level are still unknown. To fill these research gaps, this paper presents a renovated analysis of the presence of various altmetric data for scientific publications, together with a more focused discussion about the presence of altmetric data across broad subject fields and smaller research topics. The main objective of this study is two-fold: (1) to reveal the development and current situation of the presence of altmetric data across publications and subject fields, and (2) to explore the potential application of altmetric data in identifying and tracking research trends that are of interest to certain communities such as Twitter users and policy-makers. The following specific research questions are put forward:

RQ1. Compared to previous studies, how the presence of different altmetric data for WoS publications has developed until now? What is the difference of altmetric data presence across WoS publications published in different years?

RQ2. How is the presence of different altmetric data across subject fields of science? For each type of altmetric data, which subject fields show higher levels of data prevalence?

RQ3. How are the relationships among various altmetric and citation data in covering different research topics? Based on specific altmetric data, in each subject field which research topics received higher levels of altmetric attention?

**Data and methods**

*Dataset*

A total of 12,271,991 WoS papers published between 2012 and 2018 were retrieved from the CWTS in-house database. Since identifiers are necessary for matching papers with their altmetric data, only publications with a Digital Object Identifier (DOI) or a PubMed Identifier (PubMed ID) recorded in WoS were considered.

Using the two identifiers, WoS papers were matched with 12 types of altmetric data from Altmetric.com and Mendeley readership as listed in Table 1. The data from Altmetric.com were extracted from a research snapshot file with data collected up to October 2019. Mendeley readership data were separately collected through the Mendeley API in July 2019.[2] Altmetric.com provides two counting methods of altmetric performance for publications, including the number of each altmetric event that mentioned the publication and the number of unique users who mentioned the publication. To keep a parallelism with Mendeley readership, which is counted at the user level, the number of unique users was selected as the indicator for counting altmetric events in this study. For selected publications, the total number of events they accumulated on each altmetric data source are provided in Table 1 as well.

---

[2] This is to avoid the limitation in the Mendeley data reported by Altmetric.com, which is restricted to only publications with other metrics in Altmetric.com (Haustein et al., 2015).

Table 1  Descriptive statistics of 12 types of altmetric data analyzed in this study[3]

| Data source | Concept measured with regard to research outputs | NP | NE |
| --- | --- | --- | --- |
| Mendeley | Mendeley readers with the output in their Library. | 10,959,393 | 293,922,534 |
| Twitter | Twitter mentions, including public tweets, quoted tweets and retweets. | 4,173,353 | 36,092,805 |
| Facebook | Facebook mentions, including posts on a curated list of public pages only. | 1,052,235 | 2,388,875 |
| News | News media mentions on a list of news sources tracked by Altmetric.com, which contains over 5,000 English and non-English global news outlets. | 491,855 | 2,803,824 |
| Blogs | Blogs citations on a list of blogs tracked by Altmetric.com, which contains over 15,000 academic and non-academic blogs. | 448,663 | 767,381 |
| Wikipedia | Wikipedia citations on English Wikipedia pages only. | 165,170 | 239,686 |
| Policy documents | Policy document citations on a wide range of public policy documents tracked by Altmetric.com, including policy, guidance, or guidelines documents from a governmental or non-governmental organization. | 137,326 | 156,813 |
| Reddit | Reddit mentions on all sub-reddits, including original posts only. | 69,356 | 90,758 |
| F1000Prime | F1000Prime recommendations. | 69,180 | 69,197 |
| Video | Video comments on YouTube. | 48,561 | 71,191 |
| Peer review | Post-publication peer review comments collected from two forums: PubPeer and Publons. | 32,154 | 32,217 |
| Q&A | Q&A mentions on Stack Overflow. | 7,005 | 8,021 |

Note: NP refers to the number of publications with corresponding altmetric data, NE refers to the total number of corresponding altmetric events. Altmetric.com has stopped collecting data from CiteULike, Sina Weibo, LinkedIn, Pinterest, and Google+ until October 2019. Syllabus data only posted in 2015 were provided by Altmetric.com and almost all publications mentioned by Syllabus are not indexed by Web of Science. Therefore, these data sources have not been included in this study.

Besides, we collected the WoS citation counts in October 2019 for the selected publications. Citations serves as a benchmark for a better discussion and understanding of the presence and distribution of altmetric data. To keep the consistency with altmetric data, a variable citation time window from the year of publication to 2019 was utilized and self-citations were considered for our dataset of publications.

*CWTS publication-level classification system*

To study subject fields and research topics, we employed the CWTS classification system (also knowns as the *Leiden Ranking classification).* Waltman and van Eck (2012) developed this publication-level classification system mainly for citable WoS publications (Article, Review, Letter) based on their citation relations. In its 2019 version, publications are clustered into 4,535 micro-level fields of science with similar research topics (here and after known as *micro-topics*) as shown in Figure 1 with VOSviewer. For each micro-topic, the top five most characteristic terms are extracted from the titles of the publications in order to label the different micro-topics. Furthermore, these micro-topics are assigned to five main subject fields of science algorithmically obtained, including *Social Sciences and Humanities* (SSH), *Biomedical and Health Sciences* (BHS), *Physical Sciences and Engineering* (PES), *Life and Earth Sciences* (LES), and *Mathematics and Computer Science* (MCS).[4] The CWTS classification system has been applied not only in the Leiden Ranking (https://www.leidenranking.com/), but also in many different previous studies related with subject field analyses (Costas et al., 2015; Didegah & Thelwall, 2018; Zahedi & van Eck, 2018).

---

[3] See more introduction to Altmetric.com data sources at:
https://help.altmetric.com/support/solutions/articles/6000060968-what-outputs-and-sources-does-altmetric-track-
[4] See more information about CWTS classification system at: https://www.leidenranking.com/information/fields

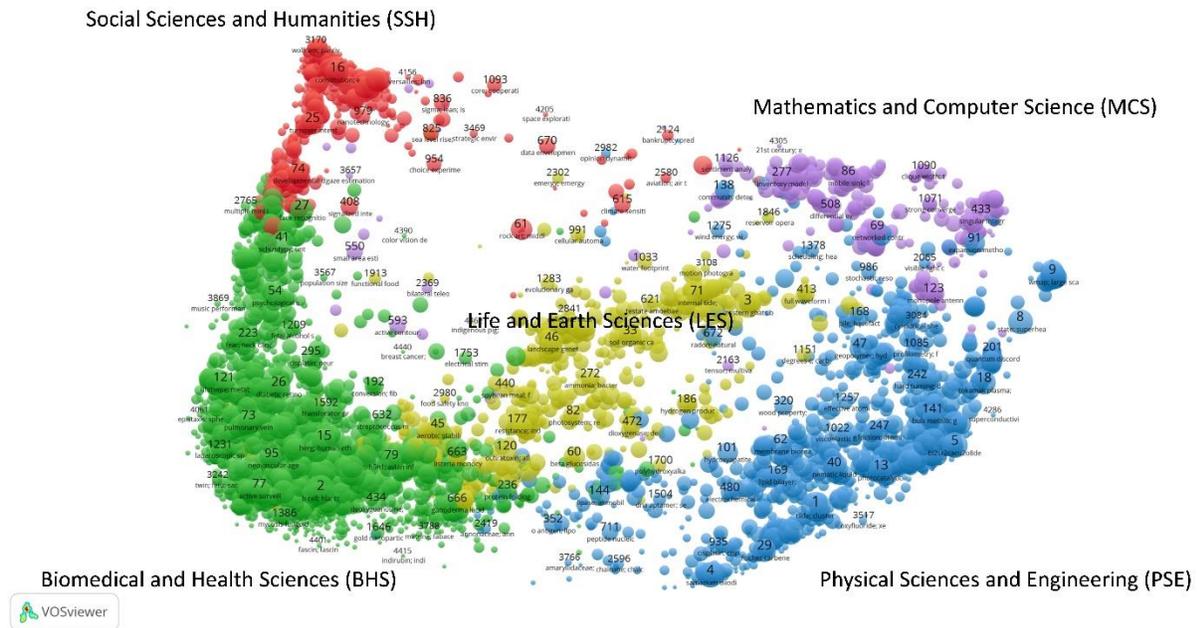

**Fig. 1** Five main subject fields of science of the CWTS classification system. Each circle represents a micro-level field (micro-topics) of clustered publications based on direct citation relations

A total of 10,615,881 of the initially selected publications (accounting for 86.5%) have CWTS classification information. This set of publications was drawn as a subset for the comparison of altmetric data presence across subject fields and research topics. Table 2 presents the number of selected publications in each main subject field.

**Table 2** Number of publications in each subject field

| Subject field | Abbr. | Number of publications | Percentage |
| --- | --- | --- | --- |
| Social Sciences and Humanities | SSH | 910,011 | 8.57% |
| Biomedical and Health Sciences | BHS | 4,272,079 | 40.24% |
| Physical Sciences and Engineering | PSE | 3,075,125 | 28.97% |
| Life and Earth Sciences | LES | 1,555,443 | 14.65% |
| Mathematics and Computer Science | MCS | 803,223 | 7.57% |

*Indicators and analytical approaches*

In order to measure the presence of different kinds of altmetric data or citation data across different sets of publications, we employed the three indicators proposed by Haustein et al. (2015): *Coverage*, *Density*, and *Intensity*. For a specific set of publications, these three indicators are defined and calculated as follows:

Coverage (C) indicates the percentage of publications with at least one altmetric event (or one citation) recorded in the set of publications. Therefore, the value of coverage ranges from 0% to 100%. The higher the coverage, the higher the share of publications with altmetric event data (or citation counts).

Density (D) is the average number of altmetric events (or citations) of the set of publications. Both publications with altmetric events (or citations) and those without any altmetric events (or citations) are considered in the calculation of density, so it is heavily influenced by the coverage and zero values.[5] The higher the value of density, the more altmetric events (or citations) received by the set of publications on average.

Intensity (I) is defined as the average number of altmetric events (or citations) of publications with at least one altmetric event (or citation) recorded. Different from D, the calculation of I only takes publications with non-zero values in each altmetric event (or citation event) into consideration, so the value must be higher or equal to one. Only in those cases of groups of publications without any altmetric events (or citations), the intensity is set to zero by default. The higher the value of intensity, the more altmetric events (or citations) that have occurred around the publications with altmetric/citation data on average.

In order to reveal the relationships among these three indicators at the research topic level, as well as the relationships of preferences for research topics among different data, the Spearman correlation analysis was performed with IBM SPSS Statistics 25.

**Results**

This section consists of four parts: the first one presents the overall presence of altmetric data for the whole set of WoS publications (in contrast with previous studies) and the evolution of altmetric data presence over the publication years. The second part compares the altmetric data presence of publications across five main subject fields of science. The third part focuses on the differences of preferences of altmetric data for research topics. In the fourth part, Twitter mentions and policy document citations are selected as two examples for identifying hot research topics with higher levels of altmetric attention received.

*Overall presence of altmetric data over the publication years*

Coverage, density, and intensity of 12 sources of altmetric data and citations were calculated for nearly 12.3 million sample WoS publications to reveal their overall presence. Table 3 presents not only the results based on our dataset, but also, for comparability purposes, the findings of data coverage (C_ref) reported by some previous altmetric empirical studies that also used Altmetric.com (and Mendeley API for Mendeley readership) as the altmetric data source, and WoS as the database for scientific publications; and also without applying restrictions of certain discipline, country, or publisher. As these previous studies analyzed datasets with size, publication years (PY), and data collection years (DY) different from ours, we present them as references for discussing the retrospective historical development of altmetric data prevalence.

---

[5] Publications without altmetric events or citations are assumed to have zero values.

**Table 3** The overall presence of 12 types of altmetric data and citation data

| Data | C | D | I | Reference | PY | DY | C_ref |
|---|---|---|---|---|---|---|---|
| Mendeley readers | 89.30% | 23.951 | 26.819 | Haustein et al. (2014) | 2010-2012 | - | 66.20% |
| | | | | Mohammadi et al. (2015) | 2008 | - | 45.60% |
| | | | | Bornmann & Haunschild (2017) | 2014 | Jul. 2016 | 89.27% |
| | | | | D'Angelo & Di Russo (2019) | 2009-2016 | Feb. 2018 | 96.10% |
| Twitter mentions | 34.01% | 2.941 | 8.648 | Robinson-García et al. (2014) | 2011-2013 | Jan. 2014 | 16.10% |
| | | | | Haustein et al. (2014) | 2010-2012 | Dec. 2012 | 9.40% |
| | | | | Costas et al. (2015) | Jul.-Dec. 2011 | Oct. 2013 | 13.30% |
| | | | | Haustein et al. (2015) | 2012 | Oct. 2013 | 21.50% |
| | | | | Meschede & Siebenlist (2018) | 2015 | - | 35.78% |
| Facebook mentions | 8.57% | 0.195 | 2.270 | Robinson-García et al. (2014) | 2011-2013 | Jan. 2014 | 3.70% |
| | | | | Costas et al. (2015) | Jul.-Dec. 2011 | Oct. 2013 | 2.50% |
| | | | | Haustein et al. (2015) | 2012 | Oct. 2013 | 4.70% |
| | | | | Meschede & Siebenlist (2018) | 2015 | - | 8.46% |
| News mentions | 4.01% | 0.229 | 5.701 | Costas et al. (2015) | Jul.-Dec. 2011 | Oct. 2013 | 0.50% |
| | | | | Haustein et al. (2015) | 2012 | Oct. 2013 | 0.70% |
| | | | | Meschede & Siebenlist (2018) | 2015 | - | 4.42% |
| Blogs citations | 3.66% | 0.063 | 1.710 | Robinson-García et al. (2014) | 2011-2013 | Jan. 2014 | 1.80% |
| | | | | Costas et al. (2015) | Jul.-Dec. 2011 | Oct. 2013 | 1.90% |
| | | | | Haustein et al. (2015) | 2012 | Oct. 2013 | 1.90% |
| | | | | Meschede & Siebenlist (2018) | 2015 | - | 2.56% |
| Wikipedia citations | 1.35% | 0.020 | 1.451 | Meschede & Siebenlist (2018) | 2015 | - | 0.70% |
| Policy document citations | 1.12% | 0.013 | 1.142 | Haunschild & Bornmann (2017) | 2000-2014 | Dec. 2015 | 0.32% |
| Reddit mentions | 0.57% | 0.007 | 1.309 | Meschede & Siebenlist (2018) | 2015 | - | 1.16% |
| F1000Prime recommendations | 0.56% | 0.006 | 1.000 | - | - | - | - |
| Video comments | 0.40% | 0.006 | 1.466 | - | - | - | - |
| Peer review comments | 0.26% | 0.003 | 1.002 | - | - | - | - |
| Q&A mentions | 0.06% | 0.001 | 1.145 | - | - | - | - |
| WoS citations | 77.43% | 9.681 | 12.502 | - | - | - | - |

According to the results, the presence of different altmetric data varies greatly. Mendeley readership provides the largest values of coverage (89.30%), density (23.95), and intensity (26.82), even higher than citations. As to other altmetric data, their presence is much lower than Mendeley readers and citations. Twitter mentions holds the second largest values among all other altmetric data, with 34.01% of publications mentioned by Twitter users and those mentioned publications accrued about 8.65 Twitter mentions on average. It is followed by several social and mainstream media data, like Facebook mentions, news mentions, and blogs citations. About 8.57% of publications have been mentioned by Facebook, 4.01% have been mentioned by news outlets, and 3.66% have been cited by blog posts. But among these three data sources, publications mentioned by news outlets accumulated more intensive attention in consideration of its higher value of intensity (5.70), which means that mentioned publications got more news mentions on average. In contrast, even though there are more publications mentioned by Facebook, they received fewer mentions at the individual publication level (with the intensity value of 2.27). For the remaining altmetric data, their data coverage values are extremely low. Wikipedia citations and policy document citations only covered respectively 1.35% and 1.12% of the sample publications, while the coverage of Reddit mentions, F1000Prime recommendations, video comments, peer review comments, and Q&A mentions are lower than 1%. In terms of these data, the altmetric data of publications are seriously zero-inflated.

Compared to the coverage reported by previous studies, an increasing trend of altmetric data presence can be observed as time goes by. Mendeley, Twitter, Facebook, news, and blogs are the most studied altmetric data sources. On the whole, the more recent the studies, the higher the values of coverage they report. Our results show one of the highest data presence for most altmetric data. Although the coverage of Twitter mentions, news mentions, and Reddit mentions reported by Meschede and Siebenlist (2018) is slightly higher than ours, it should be noted that they used a random sample consisting of 5,000 WoS papers published in 2015, and as shown in Figure 2, there exist biases toward publication years when investigating data presence for altmetrics.

After calculating the three indicators for research outputs in each publication year, Figure 2 shows the change trends of the presence of altmetric data. Overall there are two types of tendencies for all altmetric data, which are in correspondence with the accumulation velocity patterns identified in the research conducted by Fang and Costas (2020). Thus, for altmetric data with higher speed in data accumulating, such as Twitter mentions, Facebook mentions, news mentions, blogs citations, and Reddit mentions, newly published publications have higher coverage levels. In contrast, those altmetric data taking a longer time to accumulate (i.e., the slow sources defined by Fang and Costas (2020)), they tend to accumulate more prominently for older publications. Wikipedia citations, policy document citations, F1000Prime recommendations, video comments, peer review comments, and Q&A mentions fall into this "slower" category. As a matter of fact, their temporal distribution patterns resemble more that of citations counts. Regarding Mendeley readers, although it keeps quite high coverage in every publication year, it shows a downward trend as citations too, indicating a kind of readership delay, by which newly published papers have to take time to accumulate Mendeley readers (Haustein et al., 2014; Thelwall, 2017; Zahedi et al., 2017).

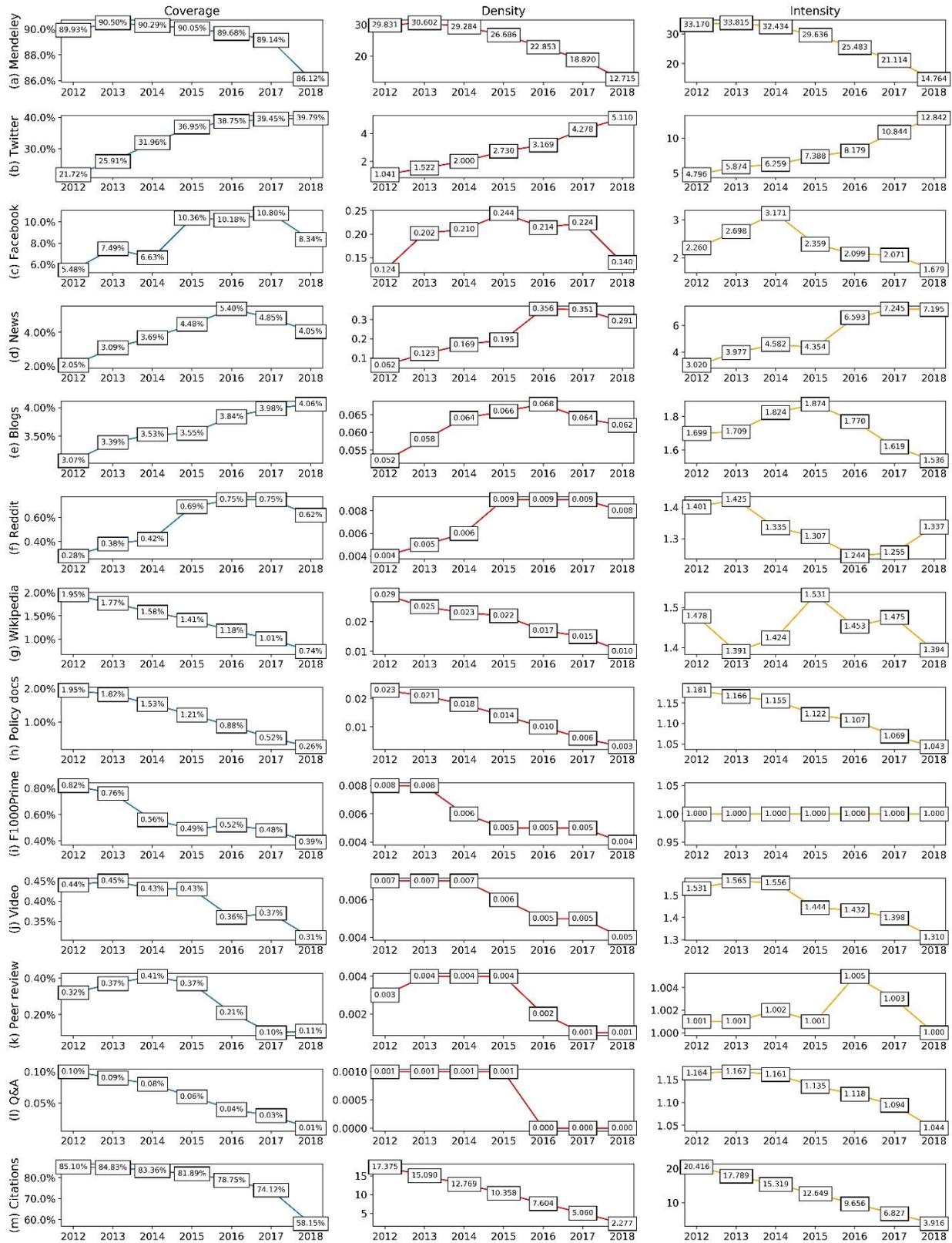

**Fig. 2** The presence of altmetric data and citations over the publication years

*Presence of altmetric data across subject fields*

In general, publications in the fields of natural sciences and medical and health sciences received more citations (Marx & Bornmann, 2015), but for altmetric data, the distribution across subject fields shows another picture. As shown in Figure 3, on the basis of our dataset, it is confirmed that publications in the subject fields of BHS, PSE, and LES hold the highest presence of citation data, and publications in the fields of SSH and MCS accumulated obviously fewer citation counts. However, as observed by Costas et al. (2015) for Twitter mentions, Facebook mentions, news mentions, blogs citations, and Google+ mentions, most altmetric data in Figure 3 are more likely to concentrate on publications from the fields of BHS, SSH, and LES, while PSE publications lose the advantage of attracting attention as they show in terms of citations, thereby performing weakly in altmetric data presence as MCS publications do.

Amongst altmetric data, there are some showing special patterns of presence. For example, PSE publications reach the coverage of Mendeley readers as high as publications in BHS, SSH, and LES, but from the perspectives of density and intensity, PSE publications drop down, showing the lowest values of density and intensity of Mendeley readers only second to MCS publications. Since F1000Prime is a platform mainly focusing on the research outputs in the fields of life sciences and medical sciences,[6] BHS publications show a considerably higher presence of F1000Prime recommendations over other subject fields. In terms of peer review comments, SSH publications hold a higher coverage level. This result differs from what has been observed in Ortega (2019)'s study on the coverage of Publons data, in which Publons data were found to be biased to publications in life sciences and health sciences. It should be noted that the peer review comment data provided by Altmetric.com is an aggregation of two platforms: Publons[7] and PubPeer[8]. In our dataset, there are 31,132 distinct publications with altmetric peer review data for the analysis of data presence across subject fields, 8,337 of them (accounting for 26.8%) having peer review comments from Publons and 22,851 of them (accounting for 73.4%) having peer review comments from PubPeer (56 publications have been commented by both). If we only consider the publications with Publons data, BHS publications and LES publications contribute the most (accounting for 53.4% and 17.2%, respectively), which is in line with Ortega (2019)'s results about Publons on the whole. Nevertheless, PubPeer data, which covers more publications recorded by Altmetric.com, is biased towards SSH publications. SSH publications make up as high as 49.9% of all publications with PubPeer data, followed by BHS publications (accounting for 43.4%), besides the relatively small quantity of WoS publications in the field of SSH, thereby leading to the overall high coverage of peer review comments of SSH publications.

Moreover, given the fact that the distributions of altmetric data are highly skewed, with the majority of publications only receiving very few altmetric events (see Appendix 1), particularly for altmetric data with relatively small data volume, their density and intensity are very close across subject fields. But in terms of intensity, there exist some remarkable subject field differences for some altmetric data. For example, on Reddit, SSH publications received more intensive attention than other subject fields in consideration of their higher value of intensity. By comparison, those LES and PSE publications cited by Wikipedia pages accumulated more intensive attention, even though the coverage of Wikipedia citations of PSE publications is rather low, suggesting that although PSE publications have a lower coverage in Wikipedia, they are more repeatedly cited.

---

[6] See more introductions to F1000Prime at: https://f1000.com/prime/faq/
[7] See more introductions to Publons at: https://publons.com/about/home/
[8] See more introductions to PubPeer at: https://pubpeer.com/

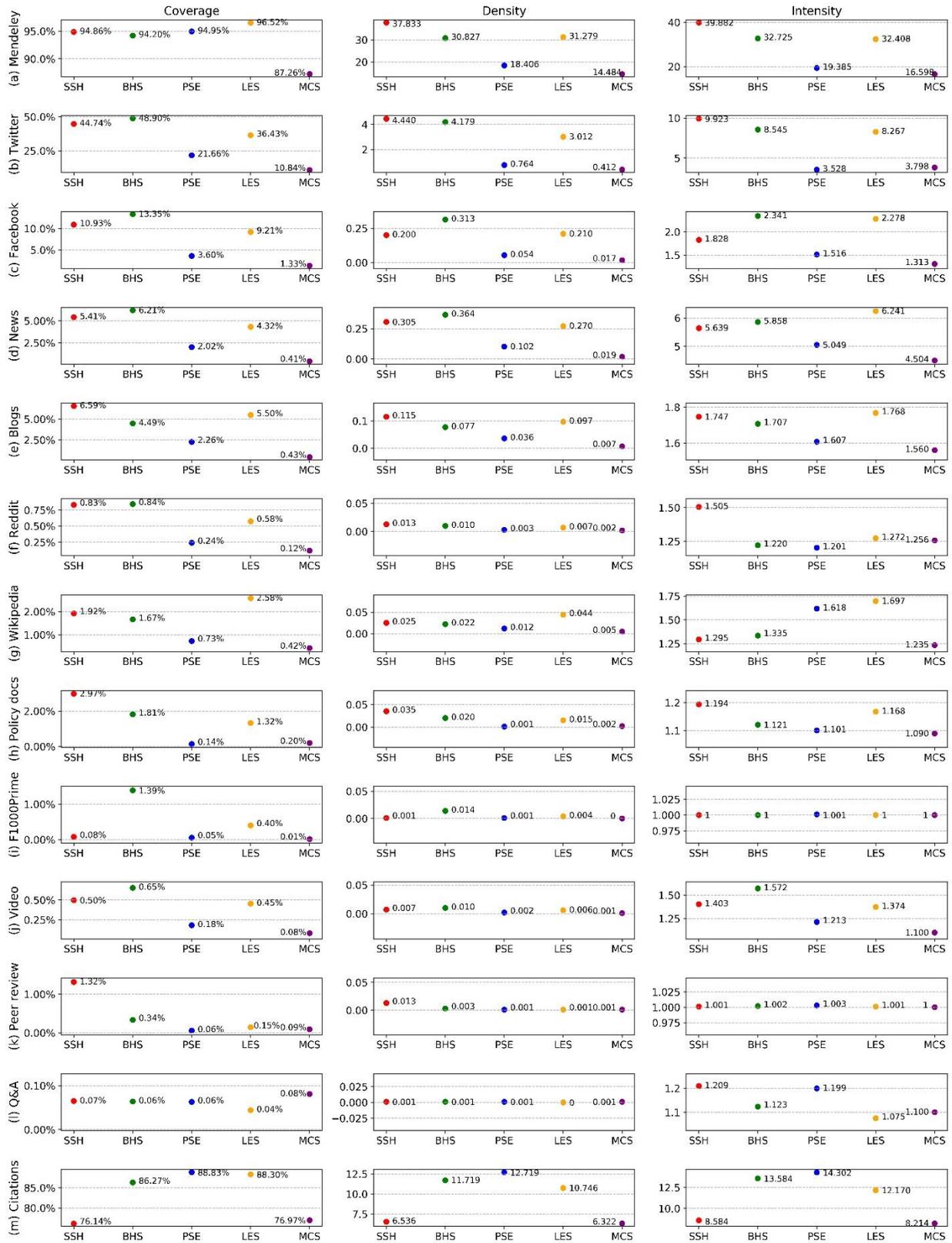

**Fig. 3** The presence of altmetric data and citations of scientific publications across five subject fields

*Presence of altmetric data across research topics*

Due to the influence of highly skewed distribution of altmetric data (see Appendix 1) on the calculation of coverage and density, these two indicators at the micro-topic level are strongly correlated for all kinds of altmetric data (see Appendix 2). In comparison, the correlation between coverage and intensity is rather weaker. Moreover, in an explicit way, coverage tells how many publications around a micro-topic have been mentioned or cited at least once, and intensity describes how frequently those publications with altmetric data or citation data have been mentioned or cited. Consequently, for a specific micro-topic, these two indicators can reflect the degree of broadness (coverage) and degree of deepness (intensity) of its received attention. Therefore, we employed coverage and intensity to investigate the presence of altmetric data at the micro-topic level and identify research topics with higher levels of attention received on different data sources.

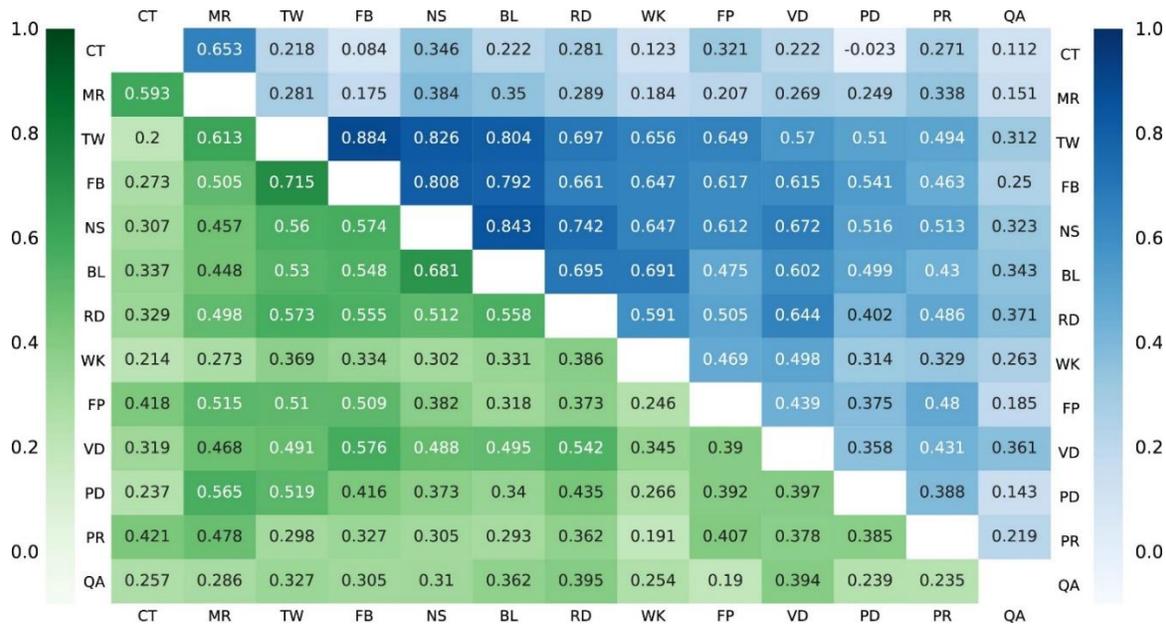

**Fig. 4** Spearman correlation analyses of coverage (upper-right triangle) and intensity (bottom-left triangle) among citations and 12 types of altmetric data at the micro-topic level. WoS citations (CT), Mendeley readers (MR), Twitter mentions (TW), Facebook mentions (FB), news mentions (NS), blogs citations (BL), Reddit mentions (RD), Wikipedia citations (WK), F1000Prime recommendations (FP), video comments (VD), policy document citations (PD), peer review comments (PR), Q&A mentions (QA).

Coverage and intensity values were calculated and appended to micro-topics based on different types of altmetric and citation data, then the Spearman correlation analyses were performed at the micro-topic level between each pair of data respectively. Figure 4 illustrates the Spearman correlations of coverage amongst citations and 12 types of altmetric data at the micro-topic level, as well as those of intensity. The higher the correlation coefficient, the more similar the presence patterns across micro-topics between two types of data. Discrepancies in the correlations can be understood as differences in the relevance of every pair of data for micro-topics, therefore some pairs of data with stronger correlations may have a more similar preference for the same micro-topics, while those with relatively weaker correlations focus on more dissimilar micro-topics. Through the lens of data coverage, Mendeley readers is the only altmetric indicator that is moderately correlated with citations at the micro-topic level, being in line with the previous conclusions about the moderate correlation between Mendeley readership counts and citations at the publication level (Zahedi et al., 2014). In contrast, because of the different distribution patterns between citations and most altmetric data across subject fields we found in Figure 3, it is not surprising that the correlations of coverage between citations and other altmetric data are relatively weak, suggesting that most altmetric data cover research topics different than citations. Among altmetric data, Twitter mentions, Facebook mentions, news mentions, and blogs citations are strongly correlated with each other, indicating that these social media data cover similar research topics. Most remaining altmetric data also present moderate correlations with the above social media data, however, Q&A mentions, as the only altmetric data showing the highest coverage of publications in the field of MCS, is weakly correlated with other altmetric data at the micro-topic level.

Nevertheless, from the perspective of intensity, most altmetric data show different attention levels towards research topics, because the values of intensity of different data are generally weakly or moderately correlated. Twitter mentions and Facebook mentions, news mentions and blogs citations, are the two pairs of altmetric data showing the strongest correlations from both coverage and intensity perspectives, thus supporting the idea that these two pairs of altmetric data do not only respectively cover very similar research topics, but also focus on similar research topics.

There exist a certain share of micro-topics in which their publications have not been mentioned at all by some specific altmetric data. In order to test the effect of those mutual zero-value micro-topics between each pair of data, the correlations have been performed also excluding them (see Appendix 3). It is observed that particularly for those pairs of altmetric data with low overall data presence across publications (e.g., Q&A mentions and peer review comments, Q&A mentions and policy document citations), their correlation coefficients are even lower when mutual zero-value micro-topics are excluded, although the overall correlation patterns across different data types at the micro-topic level are consistent with what we observed in Figure 4.

*Identification of hot research topics with altmetric data*

On the basis of coverage and intensity, it is possible to compare the altmetric data presence across research topics and to further identify topics that received higher levels of attention. As shown in Figure 5, groups of publications with similar research topics (micro-topics) can be classified into four categories according to the levels of coverage and intensity of attention received. In this framework, *hot research topics* are those topics with a high coverage level of their publications, and at the same time they have also accumulated a relatively high intensive average attention (i.e., their publications exhibit high coverage and high intensity values). Differently, those research topics in which only few publications have received relatively high intensive attention can be regarded as *star-papers topics* (i.e., low coverage and high intensity values), since the attention they attracted has not expanded to a large number of publications within the same research topic. Thus, in star-papers topics the attention is mostly concentrated around a relatively reduced set of publications, namely, those star-papers with lots of attention accrued, while most of the other publications in the same research topic do not receive attention. Following this line of reasoning, there are also research topics with a relatively large share of publications covered by a specific altmetric data, but those covered publications do not show a high average intensity of attention (i.e., high coverage and low intensity values), these research topics are defined as *popular research topics* with mile-wide and inch-deep attention accrued. Finally, *unpopular research topics* indicate those topics with few publications covered by a specific altmetric data source, and the average of data accumulated by the covered publications is also relatively small (i.e., low coverage and low intensity values); these research topics have not attracted too much attention, thereby arguably remaining in an altmetric unpopular status. It should be noted that as time goes on and with newly altmetric activity generated, the status of a research topic might switch across the above four categories.

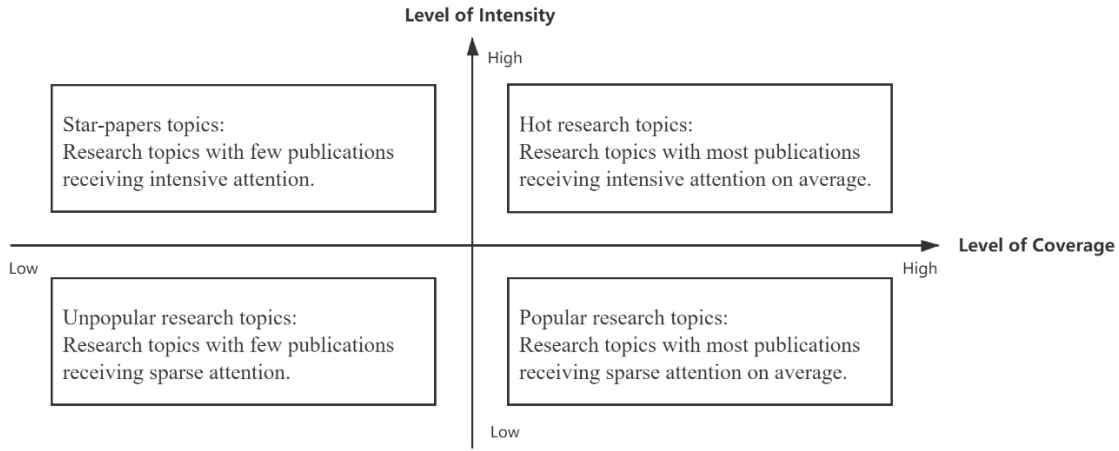

**Fig. 5** Two-dimensional system for classifying research topics with different levels of attention

Following the framework proposed in Figure 5, we took Twitter mention data as an example to empirically identify hot research topics in different subject fields. A total of 4,531 micro-topics with at least one Twitter mention in Figure 1 were plotted into a two-dimensional system according to the levels of coverage and intensity they achieved (Figure 6a). Micro-topics are ranked based on their coverage and intensity at first, respectively. The higher the ranking a micro-topic achieves, the higher the level of its coverage or intensity. Size of micro-topics is determined by their total number of publications. In order to identify representative hot research topics on Twitter, here we selected the top 10% as the criterion for both levels of coverage and intensity (two dashed lines in Figure 6a) to partition micro-topics into four parts, which are in correspondence with Figure 5. As a result, micro-topics with higher levels of coverage and intensity are classified as hot research topics that received broader and more intensive attention from Twitter users (locate at the upper right corner of Figure 6a). Because publications in the fields of SSH, BHS, and LES have much higher coverage and intensity of Twitter data, micro-topics from these three subject fields are more likely to distribute at the upper right part. In contrast, micro-topics in PSE and MCS concentrate at the lower left part. In consideration of the biased presence of Twitter data across five main subject fields, we plotted micro-topics in each subject field by the same method as Figure 6a, respectively, and then zoomed in and only presented the part of hot research topics for each subject field in Figure 6b-6f to show their identified hot research topics on Twitter. For clear visualization, one of the extracted terms by CWTS classification system was used as the label for each micro-topic.

In the field of SSH, there are 488 micro-topics considered, and 23 (5%) of them rank in top 10% from both coverage and intensity perspectives (Figure 6b). In this subject field, hot research topics tend to be about social issues, including topics related to gender and sex (e.g., "sexual orientation", "gender role conflict", "sexual harassment", etc.), education (e.g., "teacher quality", "education", "undergraduate research experience", etc.), climate ("global warming"), as well as psychological problems (e.g., "stereotype threat", "internet addiction", "stress reduction", etc.).

BHS is the biggest field with both the most research outputs and the most Twitter mentions, so there are 1,796 micro-topics considered, and 75 (4%) of them were detected as hot research topics in Figure 6c. Research topics about daily health keeping (e.g., "injury prevention", "low carbohydrate diet", "longevity", etc.), worldwide infectious diseases (e.g., "Zika virus infection", "Ebola virus", "influenza", etc.), lifestyle diseases (e.g., "obesity", "chronic neck pain", etc.), and emerging biomedical technologies (e.g., "genome editing", "telemedicine", "mobile health", etc.) received more attention on Twitter. Moreover, problems and revolutions in the medical system caused by some social activities such as "Brexit" and "public involvement" are also brought into focus.

In the field of PSE, 42 (3%) out of 1,241 micro-topics were identified as hot research topics in Figure 6d. As a field with less Twitter mentions accumulated, although most research topics are left out by Twitter users, those about the universe and astronomy (e.g., "gravitational wave", "exoplanet", "sunspot", etc.) and quantum (e.g., "quantum walk", "quantum game", "quantum gravity", etc.) received relatively higher levels of attention. In addition, there are also some hot research topics standing out from complexity sciences, such as "scale free network", "complex system", and "fluctuation theorem".

In the field of LES, there are 650 micro-topics in total, and Figure 6e shows 32 (5%) hot research topics in this field. These hot research topics are mainly about animals (e.g., "dinosauria", "shark", "dolphin", etc.) and natural environment problems (e.g., "extinction risk", "wildlife trade", "marine debris", etc.).

Finally, as the smallest subject field, MCS has 18 (5%) out of 356 micro-topics identified as hot research topics (Figure 6f), which are mainly about emerging information technologies (e.g., "big data", "virtual reality", "carsharing") and robotics (e.g., "biped robot", "uncanny valley", etc.).

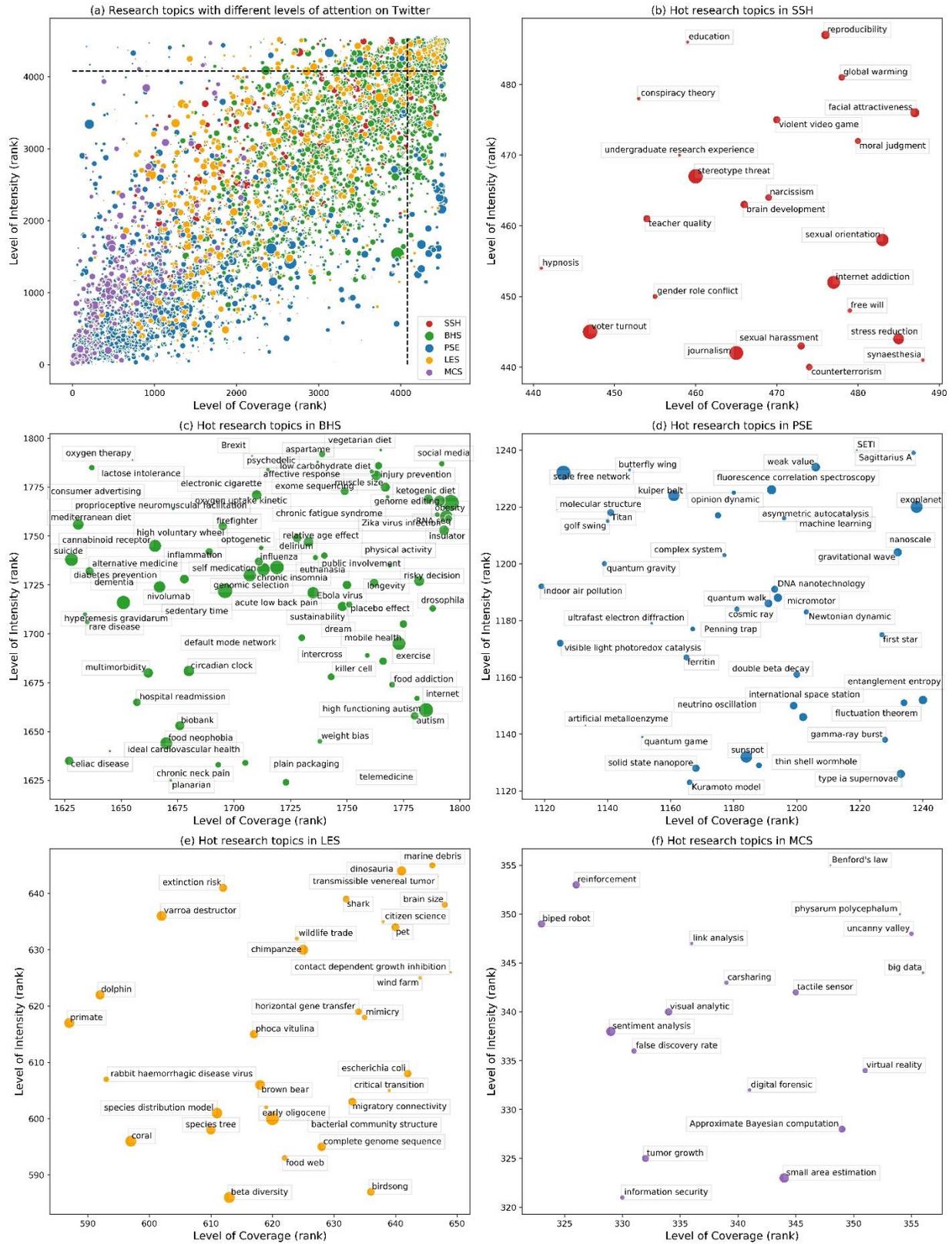

**Fig. 6** Hot research topics mentioned on Twitter in (b) SSH; (c) BHS; (d) PSE; (e) LES; (f) MCS

To reflect the differences of hot research topics through the lens of different altmetric data sources, policy document citation data was selected as another example. Figure 7 shows the overall distribution of 3,134 micro-topics with at least one policy document citation and the identified hot research topics in five main subject fields. The methodology of visualization is same as Figure 6 based on Twitter data. However, due to the smaller data volume of policy document citations, there are 1,868 micro-topics sharing the same intensity of 1. In this case, total number of policy document citations of each micro-topic was introduced as a benchmark to make distinctions. For micro-topics with the same intensity, the higher the total number of policy document citations accrued, the higher the level of attention in the dimension of intensity. After this, if micro-topics still share the same ranking, they are tied for the same place with the next equivalent rankings skipped. In general, these paralleling rankings of micro-topics with relatively low level of attention do not affect the identification of hot research topics.

Through the lens of policy document citations, identified hot research topics differ from those in the eyes of Twitter uses to some extents. In the field of SSH, 11 (3%) out of 376 micro-topics were classified as hot research topics (Figure 7b). These research topics mainly focus on industry and finance (e.g., "microfinance", "tax compliance", "intra industry trade", etc.), as well as child and education (e.g., "child care", "child labor", "teacher quality", etc.). Besides, "gender wage gap" is also a remarkable research topic appeared in policy documents.

In the field of BHS, there are 1,500 micro-topics have been cited by policy documents at least once, and 44 (3%) of them were classified as hot research topics (Figure 7c). Worldwide infectious diseases are typically concerned by policy-makers, consequently, there is no doubt that they were identified as hot research topics, such as "SARS", "Ebola virus", "Zika virus infection", and "Hepatitis C virus genotype". In addition, healthcare (e.g., "health insurance", "nursing home resident", "newborn care", etc.), social issues (e.g., "suicide", "teenage pregnancy", "food insecurity", "adolescent smoking", etc.), and potential health-threatening environment problems (e.g., "ambient air pollution", "environmental tobacco smoke", "climate change", etc.) drew high level of attention from policy-makers too.

Different from the focus of attention on astronomy of Twitter users, in the field of PSE (Figure 7d), the 16 (3%) hot research topics out of 548 micro-topics that concerned by policy-makers are mainly around energy and resources, like "energy saving", "wind energy", "hydrogen production", "shale gas reservoir", "mineral oil", and "recycled aggregate", etc.

In the field of LES, Figure 7e shows the 15 (3%) hot research topics identified out from 546 micro-topics. From the perspective of policy documents, environmental protection (e.g., "marine debris", "forest management", "sanitation", etc.) and sustainable development (e.g., "selective logging", "human activity", "agrobiodiversity", etc.) are hot research topics.

At last, in the field of MCS (Figure 7f), publications are hardly cited by policy documents, thus there are only 5 (3%) topics out of 164 micro-topics identified as hot research topics. In this field, policy-makers paid more attention to information security ("differential privacy", "sensitive question") and traffic economy ("road pricing", "carsharing").

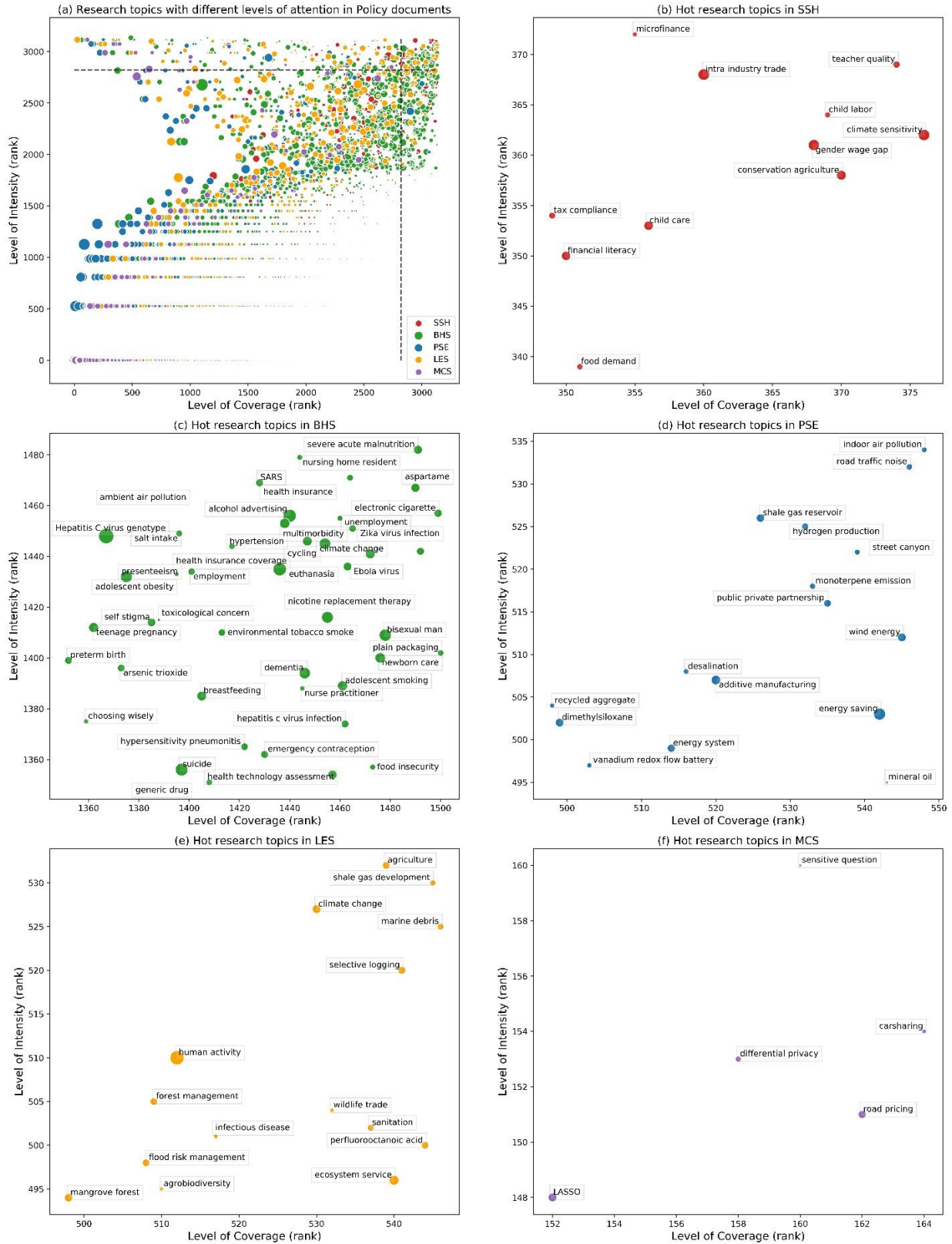

**Fig. 7** Hot research topics cited by policy documents in (b) SSH; (c) BHS; (d) PSE; (e) LES; (f) MCS

**Discussion**

*Increasing presence of altmetric data*

Data presence is essential for the application of altmetrics in research evaluation and other potential areas. The heterogeneity of altmetrics makes it difficult to establish a common conceptual framework and to draw a unified conclusion (Haustein, 2016), thus in most cases it is necessary to separate altmetrics to look into their own performance. This paper investigated 12 types of altmetric data respectively based on a large-scale and up-to-date dataset, results show that various altmetric data vary a lot in the presence for WoS publications.

Data presence of several altmetric data has been widely discussed and explored in previous studies. There are also some reviews summarizing the previous observations of the coverage of altmetric data (Erdt et al., 2016; Ortega, 2020). Generally speaking, our results confirmed the overall situations of the data presence in those studies. For instance, Mendeley readership keeps showing a very high data coverage across scientific publications and provides the most metrics among all altmetric data, followed by Twitter mentions and Facebook mentions. However, there exist huge gaps among these altmetric data. Regarding the data coverage, 89.3% of sample publications have attracted at least one Mendeley reader, while for Twitter mentions and Facebook mentions, the value is only 34.0% and 8.6%, respectively. Moreover, for those altmetric data which are hardly surveyed with the same dataset of WoS publications before, like Reddit mentions, F1000Prime recommendations, video comments, peer review comments, and Q&A mentions, their data coverage is substantially lower than 1%, showing an extremely weak data presence across research outputs.

Comparing with previous observations of altmetric data coverage reported in earlier altmetric studies, it can be concluded that the presence of altmetric data is clearly increasing, and our results are generally higher than those previous studies using the same types of datasets. There are two possible reasons for the increasing presence of altmetric data across publications. One is the progress made by altmetric data aggregators (particularly Altmetric.com), by improving their publication detection techniques and by enlarging tracked data sources. For example, Altmetric.com redeveloped their news tracking system in December 2015,[9] which partially explains the rise of news coverage in 2016 (see Figure 2). The second reason for the increasing presence of some altmetric data is the rising uptake of social media by the public, researchers, and scholarly journals (Nugroho et al., 2020; Van Noorden, 2014; Zheng et al., 2019). Against this background, scientific publications are more likely to be disseminated on social media, thereby stimulating the accumulation of altmetric data. The fact that more publications with corresponding altmetric data accrued and detected is beneficial to consolidate the data foundation, thus promoting the development and possible application of altmetrics.

In the meantime, we emphasized the biases of altmetric data towards different publication years. Costas et al. (2015) highlighted the "recent bias" they found in the overall altmetric scores, which refers to the dominance of most recent published papers in garnering altmetric data. Nevertheless, we found that the "recent bias" is not exhibited by all types of altmetric data. For altmetric data with relatively high speed in data accumulation after publication, like Twitter mentions, Facebook mentions, news mentions, blogs citations, and Reddit mentions (Fang & Costas, 2020), it is demonstrated that their temporal distribution conforms to a "recent bias". However, a "past bias" is found for altmetric data that take a relatively longer time to accumulate, such as Wikipedia citations, policy document citations, F1000Prime recommendations, video comments, peer review comments, and Q&A mentions (Fang & Costas, 2020). Due to the slower pace of these altmetric events, they are more concentrated on relatively old publications. Even for Mendeley readers, its data presence across recent publications is obviously lower.

---

[9] See more details about the data coverage date of Altmetric.com at:
https://help.altmetric.com/support/solutions/articles/6000136884-when-did-altmetric-start-tracking-attention-to-each-attention-source-

Overall, although an upward tendency of data presence has been observed over time, most altmetric data still keep an extremely low data presence, with the only exceptions of Mendeley readers and Twitter mentions. As suggested by Thelwall et al. (2013), until now these altmetric data may only be applicable to identify the occasional exceptional or above average articles rather than as universal sources of impact evidence. In addition, the distinguishing presence of altmetric data reinforces the necessity of keeping altmetrics separately in future analyses or research assessments.

*Different presence of altmetric data across subject fields and research topics*

With the information of subject fields and micro-topics assigned by the CWTS publication-level classification system, we further compared the presence of 12 types of altmetric data across subject fields of science and their inclinations to different research topics. Most altmetric data have a stronger focus on publications in the fields of SSH, BHS, and LES. In contrast, altmetric data presence in the fields of PSE and MCS are generally lower. This kind of data distribution differs from what has been observed based on citations, in what SSH are underrepresented while PSE stands out as the subject field with higher levels of citations. This finding supports the idea that altmetrics might have more added values for Social Sciences and Humanities when citations are absent (Costas et al., 2015).

In this study, it is demonstrated that even within the same subject field, altmetric data show different levels of data presence across research topics. Amongst altmetric data, their correlations at the research topic level are similar with the correlations at the publication level (Costas et al., 2015; Zahedi et al., 2014), with Mendeley readers the only altmetric data moderately correlated with citations, and Twitter mentions and Facebook mentions, news mentions and blogs citations, the two pairs showing the strongest correlations. There might exist some underlying connections within these two pairs of strongly correlated altmetric data, such as the possible synchronous updating by users who utilize multiple platforms to share science information, which can be further investigated in future research. For the remaining altmetric data, although many of them achieved moderate to strong correlations with each other from the aspect of coverage because they have similar patterns of data coverage across subject fields, the correlations of data intensity are weaker, implying that research topics garnered different levels of attention across altmetric data (Robinson-Garcia et al., 2019).

In view of the uneven distribution of specific altmetric data across research topics, it is possible to identify hot research topics which received higher levels of attention from certain communities such as Twitter users and policy-makers. Based on two indicators for measuring data presence: coverage and intensity, we developed a framework to identify hot research topics operationalized as micro-topics that fall in the first decile in terms of the ranking distribution of both coverage and intensity. This means that hot research topics are those with large shares of the publications receiving intensive average attention. We have demonstrated the application of this approach in detecting hot research topics mentioned on Twitter and cited in policy documents. Since the subject field differences are so pronounced that they might hamper generalization (Mund & Neuhäusler, 2015), the identification of hot research topics was conducted for each subject field severally. Hot research topics on Twitter reflect the interest shown by Twitter users, while those in policy documents serve as the mirror of policy-makers' focuses on science, and these two groups of identified hot research topics are diverse and hardly overlapped. This result proves that different communities are keeping an eye on different scholarly topics driven by dissimilar motivations.

The methodology of identifying hot research topics sheds light on an innovative application of altmetric data in tracking research trends with particular levels of social attention. By taking the advantage of the clustered publication sets (i.e., micro-topics) algorithmically generated by the CWTS classification system, the methodology proposed measures how wide and intensive is the altmetric attention to the research outputs of specific research topics. This approach provides a new option to monitor the focus of attention on science, thus representing an important difference with prior studies about the application of altmetric data in identifying topics of interest, which mostly were based on co-occurrence networks of topics with specific altmetric data accrued (Haunschild et al., 2019; Robinson-Garcia et al., 2019). The methodology proposed employs a two-dimensional framework to classify research topics into four main categories according to the levels of the specific altmetric attention they received. As such, the framework represents a more simplified approach to study and characterize different types of attention received by individual research topics. In our proposal for the identification of hot research topics, the influence of individual publications with extremely intensive attention received is to some extent diminished, relying the assessment of the whole topic on the overall attention of the publications around the topic, although of course those topics characterized by singularized publications with high levels of attention are also considered as "star-papers topics". It should be acknowledged that the results of this approach give an overview of the attention situations of generalized research topics, however, to get more detailed pictures of specific micro-level research fields, other complementary methods based on the detailed text information of the publications should be employed to go deep into micro-topics. Moreover, in this study, the identification of hot research topics is based on the whole dataset, in future studies, through introducing the factors of publication time of research outputs and the released time of altmetric events, it is suggested to monitor those hot research topics in real time in order to reflect the dynamic of social attention on science.

*Limitations*

There are some limitations in this study. First, the dataset of publications is restricted to publications with DOIs or PubMed IDs. The strong reliance on these identifiers is also seen as one of the challenges of altmetrics (Haustein, 2016). Second, although all types of documents are included in the overall analysis of data presence, only Article, Review, and Letter are assigned with main subject fields of science and micro-topics by the CWTS publication-level classification system, so only these three document types are considered in the following analysis of data presence across subject fields and research topics. But these three types account for 87.5% of sample publications (see Appendix 4), they can be used to reveal relatively common phenomena. Lastly, the CWTS classification system is a coarse-grained system of disciplines in consideration of that some different fields are clustered into an integral whole, like social sciences and humanities, making it difficult to present more fine-grained results. But the advantages of this system lie in that it solves the problem caused by multi-disciplinary journals, and individual publications with similar research topics are clustered into micro-level fields, namely, micro-topics, providing us with the possibility of comparing the distribution of altmetric data at the research topic level, and identifying hot research topics based on data presence.

**Conclusions**

This study investigated the state-of-the-art presence of 12 types of altmetric data for nearly 12.3 million Web of Science publications across subject fields and research topics. Except for Mendeley readers and Twitter mentions, the presence of most altmetric data is still very low, even though it is increasing over time. Altmetric data with high speed of data accumulation are biased to newly published papers, while those with lower speed bias to relatively old publications. The majority of altmetric data concentrate on publications from the fields of Biomedical and Health Sciences, Social Sciences and Humanities, and Life and Earth Sciences. These findings underline the importance of applying different altmetric data with suitable time windows and fields of science considered. Within a specific subject field, altmetric data show different preferences for research topics, thus research topics attracted different levels of attention across altmetric data sources, making it possible to identify hot research topics with higher levels of attention received in different altmetric contexts. Based on the data presence at the research topic level, a framework for identifying hot research topics with specific altmetric data was developed and applied, shedding light onto the potential of altmetric data in tracking research trends with a particular social attention focus.

**Acknowledgements**

Zhichao Fang is financially supported by the China Scholarship Council (201706060201). Rodrigo Costas is partially funded by the South African DST-NRF Centre of Excellence in Scientometrics and Science, Technology and Innovation Policy (SciSTIP). Xianwen Wang is supported by the National Natural Science Foundation of China (71673038 and 71974029). The authors thank Altmetric.com for providing the altmetric data of publications, and also thank the two anonymous reviewers for their valuable comments.

**Appendix 1**

It is reported that the distributions of citation counts (Seglen, 1992), usage counts (Wang et al., 2016), and Twitter mentions (Fang et al., 2020) are highly skewed. Results in Figure 8 show that the same situation happens to other altmetric data as well. Even though the data volume differs greatly, the distributions of all kinds of altmetric data are highly skewed, suggesting that most scientific publications only accrued few corresponding events and very few of them received high levels of attention.

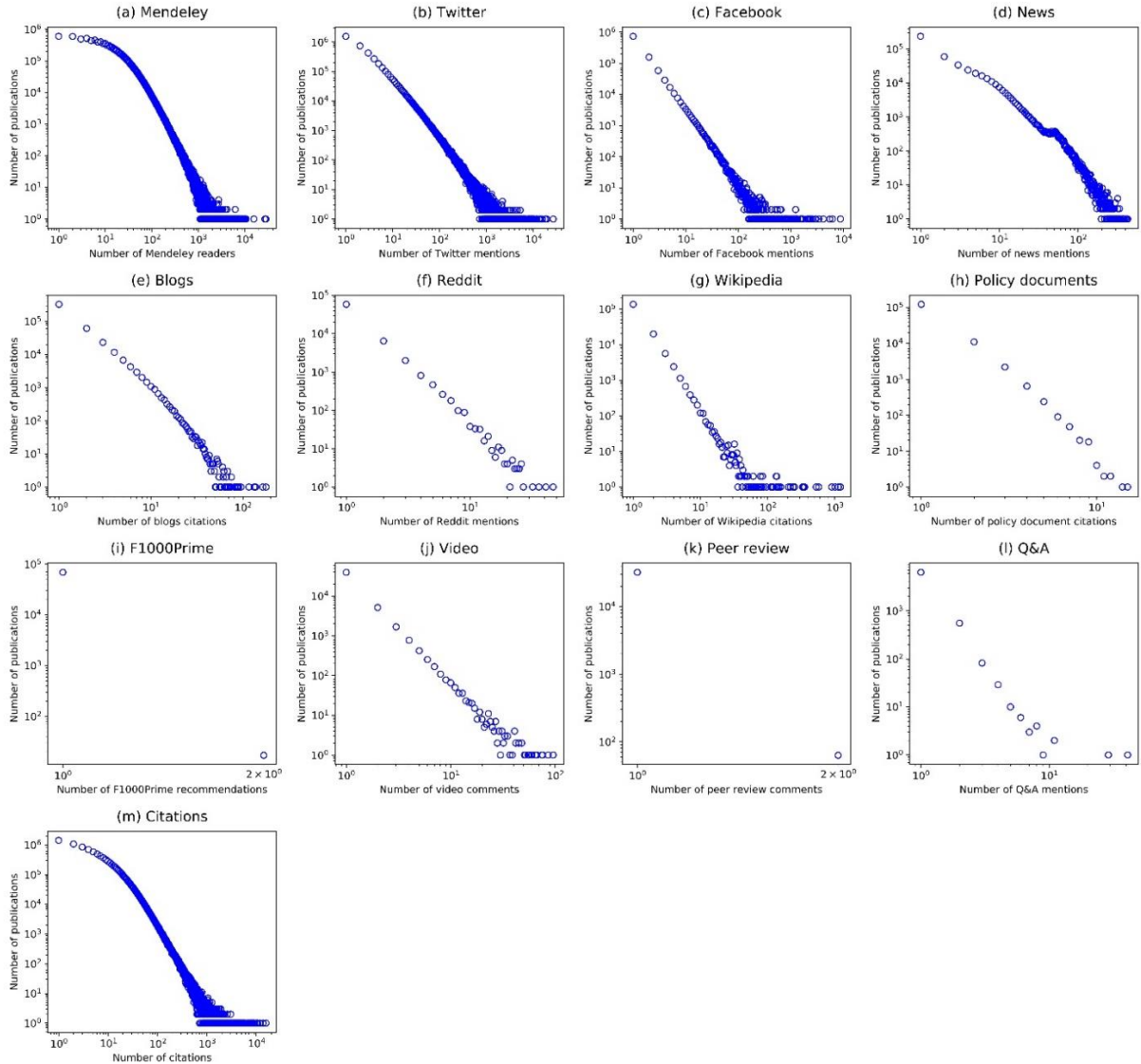

**Fig. 8** Distribution of 12 types of altmetric data and citations of sample publications

**Appendix 2**

Spearman correlation analyses among coverage, density, and intensity of micro-topics were conducted for each altmetric data and citations, and the results are shown in Figure 9. Because of the highly skewed distribution of all kinds of altmetric data, the calculation of coverage and density are prone to get similar results, especially for altmetric data with smaller data volume. Therefore, the correlation between coverage and density is quite strong for every altmetric data. For most altmetric data, density and intensity are moderately or strongly correlated, and their correlations are always slightly stronger than that between coverage and intensity.

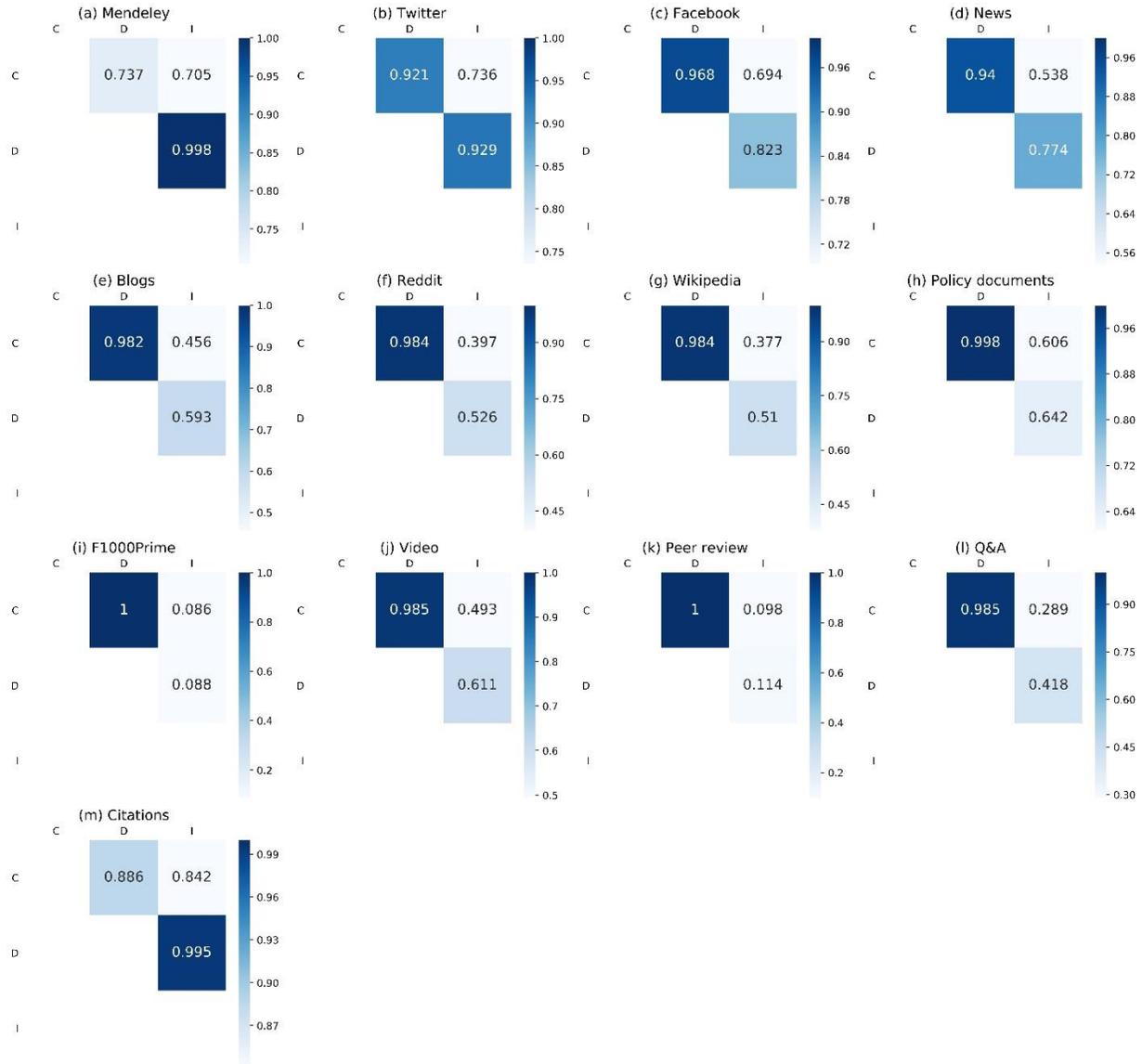

**Fig. 9** Spearman correlations among coverage (C), density (D), and intensity (I) at the micro-topic level

**Appendix 3**

In consideration of the influence of zero values of some micro-topics on inflating the Spearman correlation coefficients, we did a complementary analysis by calculating the Spearman correlations for each pair of data after excluding those mutual micro-topics with zero values (Figure 10). Compared to the results shown in Figure 4, values in Figure 10 are clearly lower, especially for those pairs of altmetric data with relatively low data presence. However, the overall patterns are still consistent with what we observed in Figure 4.

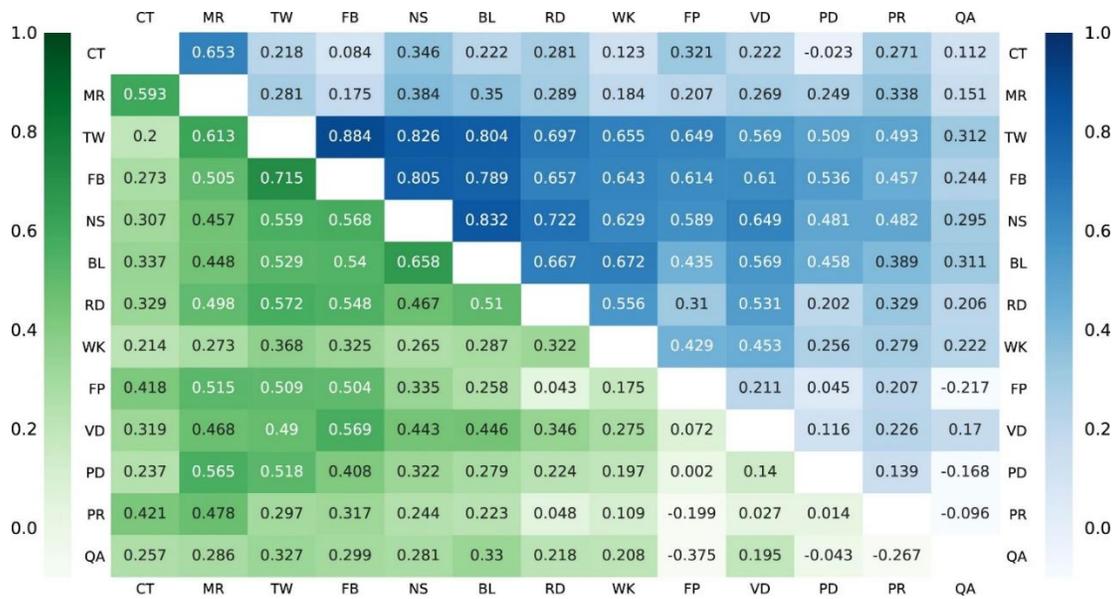

**Fig. 10** Spearman correlation analyses of coverage (upper-right triangle) and intensity (bottom-left triangle) among citations and 12 types of altmetric data at the micro-topic level (with mutual zero-value micro-topics excluded). WoS citations (CT), Mendeley readers (MR), Twitter mentions (TW), Facebook mentions (FB), news mentions (NS), blogs citations (BL), Reddit mentions (RD), Wikipedia citations (WK), F1000Prime recommendations (FP), video comments (VD), policy document citations (PD), peer review comments (PR), Q&A mentions (QA).

**Appendix 4**

The 12,271,991 sample WoS publications were matched with their document types through the CWTS in-house database. Table 4 presents the number of publications and the coverage of altmetric data of each type. The types of Article, Review, and Letter, which are included in the CWTS classification system, account for about 87.5% in total. The altmetric data coverage varies across document types as observed by Zahedi et al. (2014). For most altmetric data, Review shows the highest altmetric data coverage, followed by Article, Editorial Material, and Letter.

**Table 4** Coverage of 12 kinds of altmetric data of different document types

|  | Article | Review | Editorial Material | Meeting Abstract | Letter | Book Review | Other |
|---|---|---|---|---|---|---|---|
| Number of publications | 9,851,747 | 616,514 | 595,577 | 527,049 | 273,819 | 227,369 | 179,916 |
| Percentage | 80.28% | 5.02% | 4.85% | 4.29% | 2.23% | 1.85% | 1.47% |
| Mendeley readers | 94.27% | 95.80% | 77.02% | 46.67% | 75.02% | 31.92% | 54.99% |
| Twitter mentions | 34.61% | 55.24% | 41.74% | 2.21% | 31.72% | 10.49% | 29.09% |
| Facebook mentions | 8.30% | 16.38% | 14.97% | 0.39% | 7.79% | 2.28% | 9.03% |
| News mentions | 4.04% | 6.70% | 5.58% | 0.37% | 3.10% | 0.16% | 4.44% |
| Blogs citations | 3.75% | 6.18% | 4.52% | 0.10% | 1.86% | 0.62% | 4.04% |
| Wikipedia citations | 1.29% | 4.38% | 1.06% | 0.03% | 0.53% | 0.46% | 1.16% |
| Policy document citations | 1.16% | 2.56% | 0.90% | 0.06% | 0.53% | 0.03% | 0.33% |
| Reddit mentions | 0.56% | 0.75% | 0.81% | 0.12% | 0.38% | 0.08% | 1.38% |
| F1000Prime recommendations | 0.63% | 0.94% | 0.15% | 0.01% | 0.17% | 0.00% | 0.05% |
| Video comments | 0.39% | 1.20% | 0.35% | 0.01% | 0.16% | 0.01% | 0.27% |
| Peer review comments | 0.30% | 0.20% | 0.08% | 0.00% | 0.08% | 0.00% | 0.14% |
| Q&A mentions | 0.06% | 0.16% | 0.04% | 0.00% | 0.02% | 0.00% | 0.05% |

**References**

Alperin, J. P. (2015). Geographic variation in social media metrics: An analysis of Latin American journal articles. *Aslib Journal of Information Management*, *67*(3), 289–304. https://doi.org/10.1108/AJIM-12-2014-0176

Bornmann, L. (2014). Do altmetrics point to the broader impact of research? An overview of benefits and disadvantages of altmetrics. *Journal of Informetrics*, *8*(4), 895–903. https://doi.org/10.1016/j.joi.2014.09.005

Bornmann, L. (2015a). Usefulness of altmetrics for measuring the broader impact of research: A case study using data from PLOS and F1000Prime. *Aslib Journal of Information Management*, *67*(3), 305–319. https://doi.org/10.1108/AJIM-09-2014-0115

Bornmann, L. (2015b). Alternative metrics in scientometrics: A meta-analysis of research into three altmetrics. *Scientometrics*, *103*(3), 1123–1144. https://doi.org/10.1007/s11192-015-1565-y

Bornmann, L. (2016). What do altmetrics counts mean? A plea for content analyses. *Journal of the Association for Information Science and Technology*, *67*(4), 1016–1017. https://doi.org/10.1002/asi.23633

Bornmann, L., & Haunschild, R. (2017). Measuring field-normalized impact of papers on specific societal groups: An altmetrics study based on Mendeley Data. *Research Evaluation*, *26*(3), 230–241. https://doi.org/10.1093/reseval/rvx005

Chen, C. (2006). CiteSpace II: Detecting and visualizing emerging trends and transient patterns in scientific literature. *Journal of the American Society for Information Science and Technology*, *57*(3), 359–377. https://doi.org/10.1002/asi.20317

Costas, R., Zahedi, Z., & Wouters, P. (2015). Do "altmetrics" correlate with citations? Extensive comparison of altmetric indicators with citations from a multidisciplinary perspective. *Journal of the Association for Information Science and Technology*, *66*(10), 2003–2019. https://doi.org/10.1002/asi.23309

Crotty, D. (2014). Altmetrics: Finding meaningful needles in the data haystack. *Serials Review*, *40*(3), 141–146. https://doi.org/10.1080/00987913.2014.947839

D'Angelo, C. A., & Di Russo, S. (2019). Testing for universality of Mendeley readership distributions. *Journal of Informetrics*, *13*(2), 726–737. https://doi.org/10.1016/j.joi.2019.03.011

de Winter, J. C. F. (2015). The relationship between tweets, citations, and article views for PLOS ONE articles. *Scientometrics*, *102*(2), 1773–1779. https://doi.org/10.1007/s11192-014-1445-x

Didegah, F., & Thelwall, M. (2018). Co-saved, co-tweeted, and co-cited networks. *Journal of the Association for Information Science and Technology*, *69*(8), 959–973. https://doi.org/10.1002/asi.24028

Ding, W., & Chen, C. (2014). Dynamic topic detection and tracking: A comparison of HDP, C-word, and cocitation methods. *Journal of the Association for Information Science and Technology*, *65*(10), 2084–2097. https://doi.org/10.1002/asi.23134

Erdt, M., Nagarajan, A., Sin, S.-C. J., & Theng, Y.-L. (2016). Altmetrics: An analysis of the state-of-the-art in measuring research impact on social media. *Scientometrics*, *109*(2), 1117–1166. https://doi.org/10.1007/s11192-016-2077-0

Fang, Z., & Costas, R. (2020). Studying the accumulation velocity of altmetric data tracked by Altmetric.com. *Scientometrics*, *123*(2), 1077-1101. https://doi.org/10.1007/s11192-020-03405-9

Fang, Z., Dudek, J., & Costas, R. (2020). The stability of Twitter metrics: A study on unavailable Twitter mentions of scientific publications. *Journal of the Association for Information Science and Technology*. Advanced online publication. https://doi.org/10.1002/asi.24344

Fenner, M. (2013). What can article-level metrics do for you? *PLOS Biology*, *11*(10), e1001687. https://doi.org/10.1371/journal.pbio.1001687

Gan, C., & Wang, W. (2015). Research characteristics and status on social media in China: A bibliometric and co-word analysis. *Scientometrics*, *105*(2), 1167–1182. https://doi.org/10.1007/s11192-015-1723-2

Glänzel, W., & Czerwon, H. J. (1996). A new methodological approach to bibliographic coupling and its application to the national, regional and institutional level. *Scientometrics*, *37*(2), 195–221. https://doi.org/10.1007/BF02093621

Glänzel, Wolfgang, & Thijs, B. (2012). Using 'core documents' for detecting and labelling new emerging topics. *Scientometrics*, *91*(2), 399–416. https://doi.org/10.1007/s11192-011-0591-7

Hammarfelt, B. (2014). Using altmetrics for assessing research impact in the humanities. *Scientometrics*, *101*(2), 1419–1430. https://doi.org/10.1007/s11192-014-1261-3

Haunschild, R., & Bornmann, L. (2017). How many scientific papers are mentioned in policy-related documents? An empirical investigation using Web of Science and Altmetric data. *Scientometrics*, *110*(3), 1209–1216. https://doi.org/10.1007/s11192-016-2237-2

Haunschild, R., Leydesdorff, L., Bornmann, L., Hellsten, I., & Marx, W. (2019). Does the public discuss other topics on climate change than researchers? A comparison of explorative networks based on author keywords and hashtags. *Journal of Informetrics*, *13*(2), 695–707. https://doi.org/10.1016/j.joi.2019.03.008

Haustein, S. (2016). Grand challenges in altmetrics: Heterogeneity, data quality and dependencies. *Scientometrics*, *108*(1), 413–423. https://doi.org/10.1007/s11192-016-1910-9

Haustein, S., Bowman, T. D., & Costas, R. (2016). Interpreting 'altmetrics': Viewing acts on social media through the lens of citation and social theories. In C. R. Sugimoto (Ed.), *Theories of Informetrics and Scholarly Communication*. De Gruyter. https://doi.org/10.1515/9783110308464-022

Haustein, S., Costas, R., & Larivière, V. (2015). Characterizing social media metrics of scholarly papers: The effect of document properties and collaboration patterns. *PLOS ONE*, *10*(3), e0120495. https://doi.org/10.1371/journal.pone.0120495

Haustein, S., Larivière, V., Thelwall, M., Amyot, D., & Peters, I. (2014). Tweets vs. Mendeley readers: How do these two social media metrics differ? *It - Information Technology*, *56*(5), 207–215. https://doi.org/10.1515/itit-2014-1048

Lee, W. H. (2008). How to identify emerging research fields using scientometrics: An example in the field of Information Security. *Scientometrics*, *76*(3), 503–525. https://doi.org/10.1007/s11192-007-1898-2

Marx, W., & Bornmann, L. (2015). On the causes of subject-specific citation rates in Web of Science. *Scientometrics*, *102*(2), 1823–1827. https://doi.org/10.1007/s11192-014-1499-9

Meschede, C., & Siebenlist, T. (2018). Cross-metric compatability and inconsistencies of altmetrics. *Scientometrics*, *115*(1), 283–297. https://doi.org/10.1007/s11192-018-2674-1

Mohammadi, E., Thelwall, M., Haustein, S., & Larivière, V. (2015). Who reads research articles? An altmetrics analysis of Mendeley user categories. *Journal of the Association for Information Science and Technology*, *66*(9), 1832–1846. https://doi.org/10.1002/asi.23286

Mund, C., & Neuhäusler, P. (2015). Towards an early-stage identification of emerging topics in science—The usability of bibliometric characteristics. *Journal of Informetrics*, *9*(4), 1018–1033. https://doi.org/10.1016/j.joi.2015.09.004

Noyons, E. (2019). Measuring societal impact is as complex as ABC. *Journal of Data and Information Science*, *4*(3), 6–21. https://doi.org/10.2478/jdis-2019-0012

Nugroho, R., Paris, C., Nepal, S., Yang, J., & Zhao, W. (2020). A survey of recent methods on deriving topics from Twitter: Algorithm to evaluation. *Knowledge and Information Systems*. https://doi.org/10.1007/s10115-019-01429-z

Ortega, J. L. (2019). Exploratory analysis of Publons metrics and their relationship with bibliometric and altmetric impact. *Aslib Journal of Information Management*, *71*(1), 124–136. https://doi.org/10.1108/AJIM-06-2018-0153

Ortega, J.-L. (2020). Altmetrics data providers: A meta-analysis review of the coverage of metrics and publication. *El Profesional de La Información*, *29*(1), e290107. https://doi.org/10.3145/epi.2020.ene.07

Priem, J., Groth, P., & Taraborelli, D. (2012). The Altmetrics Collection. *PLOS ONE*, *7*(11), e48753. https://doi.org/10.1371/journal.pone.0048753

Priem, J., Taraborelli, D., Groth, P., & Neylon, C. (2010). *Altmetrics: A manifesto*. http://altmetrics.org/manifesto/. Accessed 10 Mar 2020.

Robinson-Garcia, N., Arroyo-Machado, W., & Torres-Salinas, D. (2019). Mapping social media attention in Microbiology: Identifying main topics and actors. *FEMS Microbiology Letters*, *366*(7). https://doi.org/10.1093/femsle/fnz075

Robinson-García, N., Torres-Salinas, D., Zahedi, Z., & Costas, R. (2014). New data, new possibilities: Exploring the insides of Altmetric.com. *El Profesional de La Información*, *23*(4), 359–366. https://doi.org/10.3145/epi.2014.jul.03

Seglen, P. O. (1992). The skewness of science. *Journal of the American Society for Information Science*, *43*(9), 628–638. https://doi.org/10.1002/(SICI)1097-4571(199210)43:9<628::AID-ASI5>3.0.CO;2-0

Shibata, N., Kajikawa, Y., Takeda, Y., & Matsushima, K. (2008). Detecting emerging research fronts based on topological measures in citation networks of scientific publications. *Technovation*, *28*(11), 758–775. https://doi.org/10.1016/j.technovation.2008.03.009

Small, H. (2006). Tracking and predicting growth areas in science. *Scientometrics*, *68*(3), 595–610. https://doi.org/10.1007/s11192-006-0132-y

Small, H., Boyack, K. W., & Klavans, R. (2014). Identifying emerging topics in science and technology. *Research Policy*, *43*(8), 1450–1467. https://doi.org/10.1016/j.respol.2014.02.005

Sugimoto, C. R. (2015). *"Attention is not impact" and other challenges for altmetrics*. https://www.wiley.com/network/researchers/promoting-your-article/attention-is-not-impact-and-other-challenges-for-altmetrics. Accessed 7 Mar 2020.

Sugimoto, C. R., Work, S., Larivière, V., & Haustein, S. (2017). Scholarly use of social media and altmetrics: A review of the literature. *Journal of the Association for Information Science and Technology*, *68*(9), 2037–2062. https://doi.org/10.1002/asi.23833

Thelwall, M. (2017). Are Mendeley reader counts high enough for research evaluations when articles are published? *Aslib Journal of Information Management*, *69*(2), 174–183. https://doi.org/10.1108/AJIM-01-2017-0028

Thelwall, M., Haustein, S., Larivière, V., & Sugimoto, C. R. (2013). Do altmetrics work? Twitter and ten other social web services. *PLOS ONE*, *8*(5), e64841. https://doi.org/10.1371/journal.pone.0064841

Tseng, Y.-H., Lin, Y.-I., Lee, Y.-Y., Hung, W.-C., & Lee, C.-H. (2009). A comparison of methods for detecting hot topics. *Scientometrics*, *81*(1), 73–90. https://doi.org/10.1007/s11192-009-1885-x

Van Noorden, R. (2014). Online collaboration: Scientists and the social network. *Nature News*, *512*(7513), 126. https://doi.org/10.1038/512126a

Waltman, L., & Costas, R. (2014). F1000 recommendations as a potential new data source for research evaluation: A comparison with citations. *Journal of the Association for Information Science and Technology*, *65*(3), 433–445. https://doi.org/10.1002/asi.23040

Waltman, L., & van Eck, N. J. (2012). A new methodology for constructing a publication-level classification system of science. *Journal of the American Society for Information Science and Technology*, *63*(12), 2378–2392. https://doi.org/10.1002/asi.22748

Wang, X., & Fang, Z. (2016). Detecting and tracking the real-time hot topics: A study on Computational Neuroscience. *ArXiv:1608.05517*. http://arxiv.org/abs/1608.05517

Wang, X., Fang, Z., & Sun, X. (2016). Usage patterns of scholarly articles on Web of Science: A study on Web of Science usage count. *Scientometrics*, *109*(2), 917–926. https://doi.org/10.1007/s11192-016-2093-0

Wang, X., Wang, Z., & Xu, S. (2013). Tracing scientist's research trends realtimely. *Scientometrics*, *95*(2), 717–729. https://doi.org/10.1007/s11192-012-0884-5

Wouters, P., & Costas, R. (2012). *Users, narcissism and control-tracking the impact of scholarly publications in the 21st century*. Utrecht: SURFfoundation. http://research-acumen.eu/wp-content/uploads/Users-narcissism-and-control.pdf

Wouters, P., Zahedi, Z., & Costas, R. (2019). Social media metrics for new research evaluation. In Wolfgang Glänzel, H. F. Moed, U. Schmoch, & M. Thelwall (Eds.), *Springer Handbook of Science and Technology Indicators* (pp. 687–713). Springer International Publishing. https://doi.org/10.1007/978-3-030-02511-3_26

Zahedi, Z., Costas, R., & Wouters, P. (2014). How well developed are altmetrics? A cross-disciplinary analysis of the presence of 'alternative metrics' in scientific publications. *Scientometrics*, *101*(2), 1491–1513. https://doi.org/10.1007/s11192-014-1264-0

Zahedi, Z., Costas, R., & Wouters, P. (2017). Mendeley readership as a filtering tool to identify highly cited publications. *Journal of the Association for Information Science and Technology*, *68*(10), 2511–2521. https://doi.org/10.1002/asi.23883

Zahedi, Z., & Haustein, S. (2018). On the relationships between bibliographic characteristics of scientific documents and citation and Mendeley readership counts: A large-scale analysis of Web of Science publications. *Journal of Informetrics*, *12*(1), 191–202. https://doi.org/10.1016/j.joi.2017.12.005

Zahedi, Z., & van Eck, N. J. (2018). Exploring topics of interest of Mendeley users. *Journal of Altmetrics*, *1*(1), 5. https://doi.org/10.29024/joa.7

Zheng, H., Aung, H. H., Erdt, M., Peng, T.-Q., Raamkumar, A. S., & Theng, Y.-L. (2019). Social media presence of scholarly journals. *Journal of the Association for Information Science and Technology*, *70*(3), 256–270. https://doi.org/10.1002/asi.24124